\documentclass[a4paper,11pt,twoside]{article}
\makeatletter
\usepackage{geometry,graphicx,color}
\usepackage{amsmath,amssymb}
\usepackage{hyperref}
\usepackage{enumerate}
\usepackage{dsfont}    
\numberwithin{equation}{section}
\newcommand{\institute}[1]{\newcommand{\@institute}{#1}}
\renewcommand{\maketitle}{
\vspace*{0.5\baselineskip}
{
\center\LARGE\noindent\@title\par
}%
\vspace{1.5\baselineskip}
{
\center\normalsize\noindent\ignorespaces\@author\par
}%
\vspace{0.5\baselineskip}
{
\center\normalsize\ignorespaces\@institute\par
}%
\vspace{2\baselineskip}
}%
\let\OLDthebibliography\thebibliography%
\renewcommand\thebibliography[1]{%
\OLDthebibliography{#1}%
\setlength{\parskip}{0pt}%
\setlength{\itemsep}{0pt plus 0.3ex}%
}%
\begin{document}

\title{Gauge theory models on $\kappa$-Minkowski space: Results and prospects}
\author{Kilian Hersent, Jean-Christophe Wallet}
\institute{%
\textit{IJCLab, 
Universit\'e Paris-Saclay, CNRS/IN2P3, 91405 Orsay, France
}\\%
\vskip 0,5 true cm
e-mail: {\texttt{kilian.hersent@universite-paris-saclay.fr}}, {\texttt{jean-christophe.wallet@universite-paris-saclay.fr}}\\[1ex]%
}%
\maketitle


\begin{abstract}
Recent results obtained in $\kappa$-Poincar\'e invariant gauge theories on $\kappa$-Minkowski space are reviewed and commented. A Weyl quantization procedure can be applied to convolution algebras to derive a convenient star product. For such a star product, gauge invariant polynomial action functional depending on the curvature exists only in 5 dimensions. The corresponding noncommutative differential calculus and the related connection are twisted together with the BRST structure linked to the gauge invariance.  
Phenomenological consequences stemming from the existence of one extra dimension are commented. Some consequences of the appearance of a non-vanishing one-loop tadpole upon BRST gauge-fixing are discussed.
\end{abstract}

\vfill\eject



\section{Introduction.}

\paragraph{}
Informally, the $\kappa$-Minkowski space $\mathcal{M}_\kappa$ can be viewed as the universal enveloping algebra of the Lie algebra of noncommutative coordinates\footnote{
Here, $x_0,\ x_i$ are self-adjoint operators and the deformation parameter is of order of the Planck mass: $\kappa\sim\mathcal{O}(m_{Planck})$.} 
defined by 
\begin{align}
    [x_0,x_i] &= \frac{\mathrm{i}}{\kappa}x_i, 
    & [x_i,x_j] &= 0,
    & i, j = 1, \dots, (d-1).
    \label{eq:kM_Lie_coord}
\end{align}
A more physically instructive description of $\mathcal{M}_\kappa$ is provided by its relationship to the $\kappa$-Poincar\'e algebra $\mathcal{P}_\kappa$ introduced a long time ago \cite{lukier1}. Recall that this latter involves the deformed translations and the counterpart of the rotations and boosts, these two parts acting on each other, which results in a bicrossproduct structure for $\mathcal{P}_\kappa$ \cite{majid-ruegg}. Since the $\kappa$-Minkowski space $\mathcal{M}_\kappa$ can be defined as the dual of the deformed translations, it follows that duality combined with the bicrossproduct structure result into an action of the rotations and boosts part on $\mathcal{M}_\kappa$, thus leading to the natural interpretation of $\mathcal{P}_\kappa$ as the "quantum symmetries" of $\mathcal{M}_\kappa$.

\paragraph{}
The $\kappa$-Minkowski space has been the subject of a huge literature for more than three decades now, mainly focused on algebraic properties, phenomenologically appealing proposals and various explorations of classical properties of scalar field theories built on $\mathcal{M}_\kappa$, \cite{luk2, amelin1, amelin2, habsb-impbis}. This attention is somewhat increased by the belief that $\kappa$-Minkowski space might well encode salient properties of the quantum space underlying Quantum Gravity. Compared to scalar field theories, gauge theories on $\mathcal{M}_\kappa$ are more difficult to deal with. See \cite{old-gauge1} for earlier studies and \cite{old-gauge2} for a review.

\paragraph{}
The purpose of this note is to review recent results on $\kappa$-Poincar\'e invariant gauge theory on $\mathcal{M}_\kappa$, \cite{MW20}, \cite{MW20-bis}, \cite{brst-twist}, \cite{module-paper}, \cite{tadpole}. It is organised as follows.

In section \ref{sec:star_prod}, the construction of the star product on $\kappa$-Minkowski through Weyl quantization is recalled. Results about scalar $\phi^4$ theory with such a star product are then put forward.

In section \ref{sec:gauge_th}, the construction of a gauge theory on $\kappa$-Minkowski, insisting on $\kappa$-Poincar\'e invariance, is presented. The main assumptions underlying the analysis are listed. The necessity to use twisted structures is pointed out. The differential calculus, connections, curvature are discussed. The gauge invariance of the action functional fixes the value for the dimension of the $\kappa$-Minkowski space to be $d=5$. Finally, a twisted BRST symmetry of this action is defined.

In section \ref{section4}, the above gauge theory is discussed. A closer look to the constraint $d=5$ and to some of the assumptions is taken. Phenomenological consequences are also studied together with some perturbative aspects.

Section \ref{sec:outlook} reviews approaches to gravity on noncommutative (quantum) spaces.

\section{Star-product and group algebra.}
\label{sec:star_prod}

\subsection{Convolution algebras and Weyl quantization.}\label{section21}
\paragraph{}
There is a huge literature on the subject of star products, starting from the celebrated Moyal product, generalisations of it, together with developments in the deformation theory, see e.g. \cite{Rieff, flato1, lecomte, Kons1}. It turns out that the Weyl map can be exploited to construct a star-product for algebras modeling noncommutative spaces with Lie algebra noncommutativity. In fact, this construction extends the work carried out by von Neumann \cite{vonNeum} almost one century ago aiming to define mathematical structures underlying the Weyl attempt \cite{Weyl} to settle a correspondence between classical and quantum observables, i.e. the Weyl quantization, ending up with the introduction of the Moyal product in the middle of the last century. See e.g. \cite{hennings} for a convenient presentation. One essential ingredient of the above scheme is the use of convolution algebras. The scheme can then be easily generalized, once its main steps are described by suitable mathematical objects which are synthesised below.

\paragraph{}
Given a non Abelian Lie algebra of coordinates, which corresponds to a noncommutative property of Lie algebra type in a physicist language, one starts from its corresponding Lie group $\mathcal{G}$, assumed here to be locally compact, with Haar measure $d\nu$ and modular function $\Delta$. Next, consider the convolution algebra of $\mathcal{G}$, $\mathbb{C}[\mathcal{G}]=(L^1(\mathcal{G}),\circ)$, where $\circ$ is the convolution product. Recall that $L^1(\mathcal{G})$ is the completion of the space of compactly supported continuous functions on $\mathcal{G}$ w.r.t. the norm $||F||=\int_{\mathcal{G}}d\nu(t)|F(t)|$ and can be equipped with the involution $F^*(t)=\overline{F}(t^{-1})\Delta(t^{-1})$, thus being a Banach $^*$-algebra. Here, $\overline{F}$ is the complex conjugate of $F$. 

Now, given a unitary representation of $\mathcal{G}$, $\pi_U$, with Hilbert space $\mathcal{H}$, assumed to be strongly continuous, which will be the case below, there is a non degenerate bounded representation of $\mathbb{C}[\mathcal{G}]$, $\pi:\mathbb{C}[\mathcal{G}]\to\mathcal{B}(\mathcal{H})$, defined by
\begin{equation}
   \pi(F) = \int_{\mathcal{G}} d\nu(s) F(s) \pi_U(s)
   \label{pirepst}
\end{equation}
$\mathcal{B(H)}$ denotes the set of bounded operators on $\mathcal{H}$. This representation satisfies $\pi(F\circ G)=\pi(F)\pi(G)$ and $\pi(F)^\dag=\pi(F^*)$.

Then, interpret the functions on $\mathcal{G}$ as functions in momentum space, i.e. set $F=\mathcal{F}f$ for any $F\in\mathbb{C}[\mathcal{G}]$, $\mathcal{F}$ being the Fourier transform, and use the Weyl quantization map
\begin{equation}
    Q(f) = \pi(\mathcal{F}f)
    \label{weylmap}, 
\end{equation}
which is a morphism of $^*$-algebra, $Q:\mathcal{A}\to\mathcal{B}(\mathcal{H})$, where $\mathcal{A}$ is the algebra involving the inverse Fourier transform of the above functions $F$, equipped with the star-product $\star$. Combining \eqref{weylmap} with \eqref{pirepst} yields 
\begin{align}
    f\star g &= \mathcal{F}^{-1}(\mathcal{F}f\circ\mathcal{F}g),\ 
    & f^\dag = \mathcal{F}^{-1}(\mathcal{F}(f)^*)
    \label{star-gene}.
\end{align}

\paragraph{}
As an illustration, let us apply the above general setting to present a sketchy derivation of the usual expression for the Moyal product. We thus start from $(L^1(\mathbb{H}),\circ):=\mathbb{C}[\mathbb{H}]$, the convolution algebra for the Heisenberg group $\mathbb{H}$, whose group laws are in obvious notations
\begin{align}
    g(z,u,v) g(z',u',v') & = g(z+z'+\frac{1}{2}(uv'-u'v),u+u',v+v'),\\ 
    g^{-1}(z,u,v) & = g(-z,-u,-v).
\end{align}

Recall that $\mathbb{H}$ is isomorphic to the 1-dimensional central extension of $\mathbb{R}^{2}$, with Heisenberg Lie algebra defined by the celebrated relation $[P,Q] = \mathrm{i}Z$ with $Z$ central, a one dimensional central extension of the algebra $\mathbb{R}^{2}$. The Heisenberg group $\mathbb{H}$ is unimodular (hence $\Delta=\mathds{1}$) with Haar measure given by $d\mu(g)=dzdqdp$, reducing to the Lebesgue measure. The convolution product is defined as $(F\circ G)(t)=\int_{\mathbb{H}}d\mu(s) F(s)G(s^{-1}t)$ for any $F,G\in\mathbb{C}[\mathbb{H}]$.

Next, we use the representation of $\mathbb{C}[\mathbb{H}]$,  $\pi:\mathbb{C}[\mathbb{H}]\to\mathcal{B}(L^2(\mathbb{R}))$ given by the following expression: 
\begin{equation*}
    \pi(F) = \int_{\mathbb{R}^3} dzdudv\ F(z,u,v)\pi_h[g(z,u,v)]
\end{equation*} 
where $\pi_h$ is a unitary irreducible representation\footnote{
One has $(\pi_h[g(z,u,v)]\psi)(x)=e^{i\frac{\hbar}{2} z}.e^{i(\hbar\frac{uv}{2}+vx)}\psi(x+\hbar u)$. These representations are characterized by the Stone-von Neumann theorem.} of $\mathbb{H}$. 
To get rid of the coordinate related to the central extension, we define the map $\#:L^1(\mathbb{R}^3)\to L^1(\mathbb{R}^2)$ by $F^\#(u,v):=\int dz\ F(z,u,v)e^{\mathrm{i}\frac{\hbar}{2} z}$ and thus
\begin{equation*}
    (\pi(F)\psi)(x) = \int_{\mathbb{R}^2} dudv\ F^\#(u,v)e^{\mathrm{i}(bx+\hbar\frac{pq}{2})}\psi(x+\hbar u).
\end{equation*}

Then, the action of $\#$ on the convolution product simply leads to a "twisted" convolution product, denoted by $\hat{\circ}$, now operating among functions on $\mathbb{R}^2$. From a standard computation, one gets
\begin{equation}
    (F\circ G)^\#(u,v) = \int_{\mathbb{R}^2}du'dv'\ F^\#(u',v') G^\#(u-u',v-v') e^{\frac{\mathrm{i}}{2}(uv'-u'v)},
    \label{twist-convolt}
\end{equation}
\begin{equation}
    (F\circ G)^\#(u,v) = (F^\#\hat{\circ} G^\#)(u,v). 
\end{equation}

Finally, we set $F^\#(u,v)=\mathcal{F}f(u,v)$ and apply the Weyl quantization map of \eqref{weylmap}, $Q(f):=\pi(\mathcal{F}f)$, which gives rise to $f\star g=\mathcal{F}^{-1}(\mathcal{F}f\hat{\circ}\mathcal{F}g)$, in which the twisted convolution is given by \eqref{twist-convolt}, whose expression is the usual Moyal product.

\paragraph{}
A somewhat similar derivation can be carried out to construct a star-product for deformation of the 3-dimensional space $\mathbb{R}^3_\lambda$ for which the convolution algebra for the compact group $SU(2)$ is relevant, leading to a tractable expression for this product, thanks to simplifications due to the Peter-Weil theorem which holds true here \cite{r3l-1}. Note that such a scheme could be applied as well to other "noncommutativity of Lie algebra type" and could lead to  a corresponding star-product.

We now apply the above scheme to the construction of a star-product for the $\kappa$-Minkowski space.

\subsection{Star-product for \texorpdfstring{$\kappa$}{k}-Minkowski space.}
\label{section22}
\paragraph{}
The general discussion presented in Section \ref{section21} can be applied, {\it{mutatis mutandis}}, to the case of $\kappa$-Minkowski space, generically denoted by $\mathcal{M}_\kappa$. Let us sketch below the construction. For more details, we refer to \cite{dana}, \cite{PW2018}, \cite{DS}.

\paragraph{}
One therefore starts from the Lie algebra of coordinates $\mathfrak{g}$ given by \eqref{eq:kM_Lie_coord}: $[x_0,x_i]=\frac{\mathrm{i}}{\kappa}x_i, [x_i,x_j]=0$, $i,j=1,2,...,(d-1)$, which is a solvable Lie algebra. The parameter $\kappa>0$ has mass dimension $1$. Accordingly, the related Lie group is solvable and is known to be the affine group $\mathcal{G}:=\mathbb{R}\ltimes_{{\phi}}\mathbb{R}^{d-1}$ where $\phi:\mathbb{R}\to\text{Aut}(\mathbb{R}^{d-1})$ will be given below. This locally compact group is not unimodular. The corresponding left- and right-invariant Haar measures, denoted respectively by $d\mu$ and $d\nu$, are related by ${d\nu(s)}={\Delta_{\mathcal{G}}(s)}d\mu(s),\ \forall s\in\mathcal{G}\nonumber$, where $\Delta_{\mathcal{G}}:\mathcal{G}\to\mathbb{R}^+_{/0}$ is the modular function, a continuous group homomorphism.

Next, we consider the convolution product and corresponding involution in $L^1(\mathcal{G})$ expressed in terms of the right-invariant Haar measure. The reason motivating this choice will become apparent below. The corresponding expressions read
\begin{align}
    (F\circ G)(t) &= \int_{\mathcal{G}}{d\nu(s)}\ F(ts^{-1})G(s),
    & F^*(t) &:= \overline{F}(t^{-1})\Delta_{\mathcal{G}}(t)
    \label{decadix}
\end{align}
for any $F,G\in L^1(\mathcal{G})$, where $\overline{F}$ denotes the complex conjugation of $F$. By further noticing that the group law for $\mathcal{G}$ can be expressed as
\begin{align*}
    W(p^0,\vec{p})W(q^0,\vec{q}) &= W(p^0+q^0,\vec{p}+{e^{-p^0/\kappa}\vec{q}}),
    & \text{Id}_{{\mathcal{G}}} &= W(0,0), \\ 
    W^{-1}(p^0,\vec{p}) &= W(-p^0,-e^{p^0/\kappa}\vec{p}p), &&
\end{align*}
with $\vec{p},\vec{q}\in\mathbb{R}^{d-1}$. One can write the modular function as ${\Delta_{\mathcal{G}}(p^0,\vec{p})=e^{{(d-1)}p^0/\kappa}}$ and define conveniently  $F(W)=F(p^0,\vec{p}):=\mathcal{F}f(p^0,\vec{p})$ for any $F\in L^1(\mathcal{G})$.

These expressions, combined with \eqref{decadix} give rise to
\begin{align}
    (F\circ G)(p_0,\vec{p}) &= \int_{\mathbb{ R}^d} dq_0 d^{d-1}\vec{q}\  F(p_0 - q_0, p - e^{(q_0-p_0)/\kappa} \vec{q}) G(q_0, \vec{q})
    \label{convolbis}\\
    F^*(p_0,\vec{p}) &= e^{(d-1)p_0/\kappa} \overline{F} (-p_0, -e^{p_0/\kappa}\vec{p})
    \label{involbis},
\end{align}
while it can be realized that the right-invariant Haar measure coincides with the usual Lebesgue measure, namely:
\begin{equation}
    {d\nu(W)} = dp_0 d^{d-1}\vec{p}
    \label{mesurelebesgue}.
\end{equation}

We are almost done. According to the discussion of Section \ref{section21}, we finally combine the bounded non-degenerate $^*$-representation of $L^1(\mathcal{G})$, $\pi: L^1(\mathcal{G}) \to \mathcal{B}(\mathcal{H})$, $\pi(F) = \int_{\mathcal{G}} d\nu(s) F(s) \pi_U(s)$ where $\pi_U: \mathcal{G} \to \mathcal{B}(\mathcal{H})$ is a strongly continuous unitary representation of $\mathcal{G}$, with the Weyl quantization map $Q$ defined by \eqref{weylmap}: $Q(f)=\pi(\mathcal{F}f)$. It satisfies
\begin{align}
    Q(f\star g) &= Q(f) Q(g), & 
    (Q(f))^* &= Q(f^\dag)
    \label{propQ},
\end{align}
with
\begin{align}
    f\star g &= \mathcal{F}^{-1}(\mathcal{F}f\circ\mathcal{F}g), &
    f^\dag &= \mathcal{F}^{-1}(\mathcal{F}(f)^*)
    \label{starinvol-def},
\end{align}
inherited from
\begin{align}
    \pi(F\circ G) &= \pi(F) \pi(G), &
    \pi(F)^\dag &= \pi(F^*)
    \label{basic-prop}.
\end{align}
One obtains after a standard calculation:
\begin{align}
    (f\star g)(x) &= \int \frac{dp^0}{2\pi} dy_0\ e^{-\mathrm{i}y_0p^0} f(x_0+y_0, \vec{x}) g(x_0, e^{-p^0/\kappa}\vec{x}), 
    \label{star-kappa}\\
    f^\dag(x) &= \int \frac{dp^0}{2\pi} dy_0\ e^{-\mathrm{i}y_0p^0} \overline{f} (x_0+y_0, e^{-p^0/\kappa} \vec{x})
    \label{invol-kappa},
\end{align}
which thus define the star-product for $\kappa$-Minkowski space and the corresponding involution.

\paragraph{}
Some comments are in order.

First, it is clear from the above construction that the star-product \eqref{star-kappa} can be viewed as the Fourier transform of the convolution product related to the affine group $\mathcal{G}$. Besides, it turns out that the star-product does not depend on the representation, which stems merely from the non-degeneracy of the representation $\pi$.

Next, we note that \eqref{star-kappa} and \eqref{invol-kappa} extend to a suitable multiplier algebra involving coordinate functions, constants, etc ..., which is convenient for physical applications. This has been nicely done in \cite{DS}. As mentioned above, $\mathcal{G}$ is a solvable group. Hence it is amenable. It follows that the corresponding group $C^*$-algebra is simply isomorphic to the reduced $C^*$-algebra which is merely obtained by completing $(L^1(\mathcal{G}),\circ)$ w.r.t. the left regular representation on $L^2(\mathcal{G})$. From general results on $C^*$-algebras, the above $C^*$-algebra involves a dense subalgebra made of functions with compact support in $p_0$ taking values in the set of smooth functions in $p_1$ with compact support. From a mere Fourier transform, one moves to the corresponding set of functions depending on the space-time variables. The algebra of functions is in particular analytic and of exponential form in $x_0$, \textit{i.e.} such that the corresponding analytic continuation is an entire function on $\mathbb{C}$ with an exponential bound.

Finally, it is worth stressing that the above star-product proves to be very convenient in practical applications to noncommutative field theories defined on $\kappa$-Minkowski spaces. Some corresponding results will be briefly reviewed in Section \ref{section23}.

\paragraph{}
From now on, we denote generically the multiplier algebra by $\mathcal{M}_\kappa$.

It turns out that the Lebesgue integral (i.e. the right-invariant measure) defines a twisted trace w.r.t. the star-product \eqref{star-kappa}, which is expressed by the following formula
\begin{equation}
    \int d^dx\ (f\star g)(x) = \int d^dx\ ((\mathcal{E}^{d-1}\triangleright g) \star  f)(x),
    \label{twisted-trace}
\end{equation}
for any $f,g \in \mathcal{M}_\kappa$ where the set $\{\mathcal{E}=e^{-P_0/\kappa}, P_1, \dots, P_d\}$ generates the Hopf-subalgebra $\mathcal{T}_\kappa$ of the Hopf $\kappa$-Poincar\'e algebra $\mathcal{P}_\kappa$ in the notations of ref. \cite{PW2018}. Hence the cyclicity of the trace w.r.t. the start-product is lost. 

But not all is lost! Cyclicity is traded for a new property. Indeed any functional of the form $\varphi(f)=\int d^dx\ f$ defines a KMS weight \cite{kuster} for the modular group of automorphisms $\sigma_t(f)=e^{it(d-1)P_0/\kappa}\triangleright f$. For a discussion on the link with the Tomita-Takesaki modular theory \cite{takesaki} and possible physical implications including the possibility to define a global time observer independant, see \cite{PW2018}. In the following, we will call "modular twist" the automorphism defined by $\mathcal{E}^{d-1}\triangleright$ in eqn.\eqref{twisted-trace}. Observe that it depends on the dimension $d$ of the $\kappa$-Minkowski space.

It appears that the above KMS weight is invariant under the action of the $\kappa$-Poincar\'e algebra $\mathcal{P}_\kappa$. This translates into \cite{DS}
\begin{equation}
    h\blacktriangleright S 
    = \int d^dx\ h\triangleright\mathcal{L} 
    = \epsilon(h)S
    \label{poinca-invar},
\end{equation}
for any $h\in\mathcal{P}_\kappa$ where $\epsilon$ is the co-unit of $\mathcal{P}_\kappa$. This last relation implies that any action functional involving the Lebesgue integral is $\kappa$-Poincar\'e invariant.

\paragraph{}
As far as field theories on $\kappa$-Minkowski space are concerned, requiring $\kappa$-Poincar\'e invariance of the corresponding action functional appears to be a physically natural requirement. Indeed, it amounts to assume that the $\kappa$-Poincar\'e symmmetry is the relevant symmetry of the quantum space valid at an energy scale close to $\kappa$, traditionally assumed to be close to the Planck mass, which reduces to the usual Poincar\'e symmetry at low energy scale for which the Minkowski space becomes relevant.

\subsection{Past (one-loop) results for noncommutative scalar fields theories on \texorpdfstring{$\kappa$}{k}-Minkowski}
\label{section23}
\paragraph{}
We denote by $\langle.,.\rangle$ the Hilbert product on $\mathcal{M}_\kappa$ defined by
\begin{equation}
    \langle f,g \rangle := \int d^dx\ (f^\dag\star g)(x)
    \label{hilbert-prod}
\end{equation}
for any $f,g\in\mathbb{A}$.

\paragraph{}
The above star-product, defining a deformation of the Minkowski space, has been recently used to explore the one-loop properties of different classes of $\kappa$-Poincar\'e-invariant noncommutative field theories on $\mathcal{M}_\kappa$ in 4 (engineering) dimensions, characterized: i) by various kinetic terms $\sim$ $\langle \phi, K_\kappa\phi \rangle,\ \ \langle \phi^\dag,K_\kappa\phi^\dag \rangle$ with kinetic operators $K_\kappa(\mathcal{C}_\kappa)$ depending on the Casimir operator of $\mathcal{P}_\kappa$, namely $\mathcal{C}_\kappa(P_\mu) = 4\kappa^2 \sinh^2\left(\frac{P_0}{2\kappa}\right) + e^{P_0/\kappa} \vec{P}^2 $, and ii) for different type of ($\phi^4$-type) interactions. These latter can be classified, according to the liturgy of noncommutative field theories as orientable interactions and non-orientable interaction, respectively of the form
\begin{align}
  \langle \phi^\dag \star \phi,& \phi^\dag \star \phi\rangle, & 
  \langle \phi \star \phi^\dag,& \phi \star \phi^\dag \rangle,
  \label{oriantab} \\
  \langle \phi^\dag \star \phi^\dag,& \phi^\dag \star \phi^\dag \rangle, & 
  \langle \phi \star \phi,& \phi \star \phi \rangle.
  \label{nonorientab} 
\end{align}
It appears that the computations of the loop diagrams can be rather easily performed, thanks to the simple structure of the star-product \eqref{star-kappa}. We refer to \cite{PW2018}, \cite{PW2018bis}, \cite{Juric:2018qdi} for technical details.

\paragraph{}
Let us synthesize the results obtained so far. First, it appears that the modular twist, which shows up in the course of the computations due to the presence of the twisted trace \eqref{twisted-trace}, controls partially the UV behaviour of the various diagrams. It generates, in particular, different new UV behaviour among the planar and non-planar diagrams, thus resulting in new sub-types of diagrams. Nevertheless, the overall (somewhat rough) observation is that the UV behaviour of the diagrams stays "close" to the one of the usual complex $\phi^4$ 4-dimensional model.

\paragraph{}
Including results on the UV/IR mixing which may potentially occur in noncommutative field theories, producing a detrimental effect on the perturbative renormalisability, yields the following overall conclusions:
\begin{itemize}
    \item Noncommutative complex scalar models involving orientable interactions, namely of the type  $\int d^4x\ \phi^\dag\star\phi\star\phi^\dag\star\phi$, do not suffer from UV/IR mixing. Two-point functions do not exhibit IR singularities responsible for the occurrence of mixing, while for "reasonable" kinetic operators, these two-point functions exhibit at most linear UV divergences.
    \item Noncommutative complex scalar models involving non-orientable interactions of the type $\int d^4x\ \phi\star\phi\star\phi^\dag\star\phi^\dag$ are plagged by UV/IR mixing, as expected. This is clearly signaled by those specific IR singularities in the two-point functions, which generates the mixing. Note that the UV behaviour of these two-point functions is somewhat similar to the one encountered for orientable interactions, except whenever the kinetic operator coincides with the Casimir operator for which quadratic UV divergences can occur.
\end{itemize}

Notice that one interesting noncommutative field theories with orientable interactions has been considered in \cite{PW2018bis}. It is characterised by a kinetic operator given by $K^{eq}_\kappa(P_\mu)=D^{eq}_0D^{eq}_0+\sum_i D^{eq}_iD^{eq}_i$ where
\begin{align}
    D^{eq}_0 &:= \frac{\mathcal{E}^{-1}}{2} \left( \kappa(1-\mathcal{E}^2) - \frac{1}{\kappa} \vec{P}^2 \right), &
    D^{eq}_i &:= \mathcal{E}^{-1}P_i,
    \label{francesco}
\end{align}
is the equivariant Dirac operator considered in \cite{francesco1}. For this field theory, the one-loop beta function vanishes \cite{PW2018bis} which signals the scale-invariance of the coupling constant, at least at this order.

\paragraph{}
The next logical goal in this exploration is to examine if $\kappa$-Poincar\'e invariance can be accommodated with gauge invariance. This is not so straightforward and as a first step, we will consider noncommutative analogs of the celebrated Yang-Mills theories which technically means that we will consider noncommutative connections on a right module over the algebra modeling $\mathcal{M}_\kappa$. This step is an essential preliminary before examining extensions involving   noncommutative analogs of affine and Levi-Civita connections.

It appears that $\kappa$-Poincar\'e-invariant gauge theories on $\mathcal{M}_\kappa$ can be obtained but are rather strongly constrained, as we now show.

\section{Gauge theory models on \texorpdfstring{$\kappa$}{k}-Minkowski space.}
\label{sec:gauge_th}

\subsection{Gauge invariance and twisted trace.}
\label{section31}
\paragraph{}
There are a few assumptions underlying the construction of $\kappa$-Poincar\'e-invariant gauge theories on $\kappa$-Minkowski spaces \cite{MW20}, \cite{MW20-bis}. These assumptions are explicitly listed below, the two first ones stemming from physical requirements while the three additional assumptions are technical and are of course needed in order to obtain explicit formulas for an invariant action functional.

The physics oriented assumptions are:
\begin{itemize}
    \item \label{it:a1}
    The action denoted by $S_\kappa$ is a polynomial functional depending on the curvature and is both invariant under $\mathcal{P}_\kappa$ and the noncommutative gauge symmetry, assumes for simplicity to be a noncommutative analog of a $U(1)$ symmetry,
    \item \label{it:a2}
    The commutative limit of $S_\kappa$, i.e. the limit $\kappa\to\infty$, coincides with the action describing an ordinary gauge theory. 
\end{itemize}
The three technical assumptions supplementing the basic structural assumptions listed above are:
\begin{itemize}
    \item \label{it:a3}
    We assume that the twisted versions of noncommutative connection are defined on a right module $\mathbb{E}$ over $\mathcal{M}_\kappa$. This is suitable for our purpose of constructing noncommutative analogs of Yang-Mills theories.
    \item \label{it:a4} 
    The module $\mathbb{E}$  is assumed to be a copy of $\mathcal{M}_\kappa$, i.e. $\mathbb{E}\simeq\mathcal{M}_\kappa$, equipped with an hermitean structure. This permits one to model noncommutative analog of the $U(1)$ gauge symmetry.
    \item \label{it:a5}
    We assume that the action of $\mathcal{M}_\kappa$ on $\mathbb{E}$, defined as a linear map $\Phi:\mathbb{E}\otimes\mathcal{M}_\kappa\to\mathbb{E}$, is twisted by an automorphim of $\mathcal{M}_\kappa$. Namely, one has:
    \begin{align}
        \Phi(m\otimes f) &:= m \bullet f = m \star \sigma(f), & \sigma &\in \mathrm{Aut}(\mathcal{M}_\kappa)
    \label{zeaction},
    \end{align}
    for any $m\in\mathbb{E}$ and $f\in\mathcal{M}_\kappa$.
\end{itemize}
For the moment, we will restrict ourselves to 
\begin{equation*}
   \sigma=\text{Id},  
\end{equation*}
thus considering
untwisted action of the algebra on the module as in \cite{MW20}, i.e.
\begin{equation}
   m\bullet f=m\star f.
\end{equation}

Besides, $\kappa$-Poincar\'e invariance of the action functional is insured for any functional of the form occurring in \eqref{poinca-invar}, a form which will be assumed from now on.

\paragraph{}
An immediate problem arises due to the loss of cyclicity of the trace. Indeed, using the popular description of a noncommutative connections on a right-module (see \ref{it:a4}) and using a standard (untwisted) noncommutative differential calculus, some by now standard considerations give rise to a curvature with the following gauge transformations
\begin{equation}
    F_{\mu\nu}^g = g^\dag \star F_{\mu\nu} \star g,
    \label{untwis-gauge}
\end{equation}
for any $g\in\mathcal{U}$, the gauge group generated by the unitary elements of the module $\mathbb{E}$, namely
\begin{equation}
    \mathcal{U}=\{g\in\mathbb{E},\ g^\dag\star g=g\star g^\dag= \mathds{1} \}, 
    \label{gauggroup}
\end{equation}
which can be viewed as the noncommutative analog of $U(1)$. Then, looking for a gauge invariant polynomial in the curvature fullfilling \ref{it:a1} and \ref{it:a2}, one is naturally led to consider e.g. the gauge invariance of $\langle F,F\rangle =\int d^dx\  F_{\mu\nu}^\dag\star F_{\mu\nu}$. But standard manipulations yield
\begin{equation}
    \langle F^g,F^g\rangle = \int d^dx\ {(\mathcal{E}^{d-1}(g)\star g^\dag)} \star F_{\mu\nu}^\dag\star F_{\mu\nu}
    \label{zepb}.
\end{equation}
One concludes that 
\begin{equation}
  \langle F^g,F^g\rangle\ne \langle F,F\rangle,
\end{equation}
signaling the lack of gauge invariance.

\paragraph{}
One way out would be to impose $\mathcal{E}^{d-1}(g)\star g^\dag=\mathds{1}$ which amounts to restrict {\it{formally}} the gauge group to 
\begin{equation}
    \mathcal{U}_\Delta = \{ g\in\mathcal{U}, \mathcal{E}^{d-1}(g)=g \} \subset \mathcal{U}. 
\end{equation}
However, the condition $\mathcal{E}^{d-1}(g)=g$ imposes very (actually too) strong restrictions on the possible functions $g$. In fact, according to the discussion in Section \ref{section22}, elements of $\mathbb{E} \simeq \mathcal{M}_\kappa$ admit an analytic continuation in $x_0$ which is an entire function. Now from the action of $\mathcal{E}$ given by 
\begin{equation}
 (\mathcal{E}\triangleright g)(x) 
 = (e^{-P_0/\kappa}\triangleright g)(x) 
 = g(x_0 + \frac{\mathrm{i}}{\kappa}, \vec{x}),   
\end{equation}
one infers that $g$ is periodic. But by the Liouville theorem, any entire periodic function is constant.

\paragraph{}
Another way to escape \eqref{zepb}, thus obtaining a gauge invariant action functional, would be the use of a suitably twisted differential, together with a related notion of twisted connection and curvature, leading to a twisted version of \eqref{untwis-gauge} able to compensate the factor $\mathcal{E}^{d-1}(g)\star g^\dag$ in \eqref{zepb}.\\
Twisted structures already appeared in twisted spectral triples which are naturally related to twisted differential calculi. Twisted spectral triples have been used in various contexts, see e.g. \cite{como-1}, \cite{martinet}, \cite{Devastato-Martinetti, Filaci-Martinetti}. In short, a twisted spectral triple with Dirac operator $D$ satisfies the condition that $[D,f]_\rho := Df-\rho(f)D$, for any $f$ in the algebra $\mathcal{A}$ of the triple, is a bounded operator on the Hilbert space of the triple. Here, $\rho \in \mathrm{Aut}(\mathcal{A})$ is the twist which is a regular automorphism, namely
\begin{equation}
    \rho(f)^\dag = \rho^{-1}(f^\dag)
    \label{regular}
\end{equation}
for any $f\in\mathcal{A}$. We will come back to this last property later on in connection with the present construction of a gauge theory on $\kappa$-Minkowski space. It appears that $[D,f]_\rho$ defined above acts as a twisted derivation on $\mathcal{A}$. Indeed, one has (product of algebra understood)
\begin{equation}
    \delta_\rho(fk)
    := [D,fk]_\rho 
    = [D,f]_\rho k + \rho(f)[D,k]_\rho 
    = \delta_\rho(f) k + \rho(f) \delta_\rho(k)
\end{equation}
for any $f,k\in\mathcal{A}$. It can be further extended to a derivation of $\mathcal{A}$ in the set of 1-forms defined by $\Omega^1_D=\{\omega=\sum_i f_i[D,k_i],\ f_i,k_i\in\mathbb{A} \}$, up to technical conditions to be satisfied\footnote{
$\Omega^1_D$ must be a $\mathcal{A}-\mathcal{A}$-bimodule with action $f\bullet\omega\bullet k=\rho(f)\omega k$, for any $f,k\in\mathcal{A}$.}.
 
\paragraph{}
In the present situation, one convenient way to obtain twisted differential calculi is to make use of twisted derivations to build derivation-based differential calculi \cite{mdv}. Recall that a twisted derivation is a linear map $X:\mathcal{A}\to\mathcal{A}$ satisfying a twisted Leibnitz rule
\begin{equation}
    X(a\star b) = X(a)\star \alpha(b) + \beta(a)\star X_(b),
    \label{gene-twist-leib}
\end{equation}
for any $a,b\in\mathcal{A}$, where $\mathcal{A}$ will be identified in a short while with the algebra modeling the $\kappa$-Minkowski space and $\star$ with the corresponding star-product. In \eqref{gene-twist-leib}, $\alpha, \beta \in \mathrm{Aut}(\mathcal{A})$. Twisted derivations have been considered from various viewpoints in algebra, see e.g. \cite{ore}, \cite{Hom-Lie}. It appears that they can be used to build twisted versions of the standard (untwisted) derivation-based differential calculi \cite{mdv, der-based, jcw-gauge2} as it will be summarized in the next subsection.

\subsection{Twisted differential calculus.}
\paragraph{}
In view of the duality between deformed translations $\mathcal{T}_\kappa\subset\mathcal{P}_\kappa$ and $\mathcal{M}_\kappa$ together with the assumption \ref{it:a2} given at the beginning of Section \ref{section31}, it is natural to look for twisted derivations as elements in $\mathcal{T}_\kappa$.\footnote{
Recall that $\mathcal{M}_\kappa$ is a left-Hopf algebra module over $\mathcal{T}_\kappa$. The corresponding action is $(\mathcal{E}\triangleright f)(x) = (e^{-P_0/\kappa}\triangleright f)(x) = f(x_0+\frac{\mathrm{i}}{\kappa},\vec{x}) $, $(P_\mu\triangleright f)(x) = -\mathrm{i}(\partial_\mu f)(x)$.}

As shown in detail in \cite{MW20}, \cite{MW20-bis}, there exists a unique Abelian Lie  algebra of twisted derivations of the Hopf subalgebra $\mathcal{T}_\kappa$ defined by (in the notations of \cite{MW20}, \cite{MW20-bis}):
\begin{equation}
    \mathfrak{D}_\gamma=\big\{X_\mu:\mathcal{M}_\kappa\to \mathcal{M}_\kappa,\ \  X_0=\kappa\mathcal{E}^\gamma(1-\mathcal{E}),\ \ X_i=\mathcal{E}^\gamma P_i,\ \  i=1,2,...,d-1\big\}
\label{tausig-famil},
\end{equation}
where $\gamma\in\mathbb{R}$ and $d$ denotes the engineering dimension of the $\kappa$-Minkowski space, to be entirely fixed in a while. $\mathfrak{D}_\gamma$ satisfies $[X_\mu,X_\nu]:=X_\mu X_\nu-X_\nu X_\mu=0$ and
\begin{equation}
    X_\mu(f\star h) = X_\mu(f) \star \mathcal{E}^\gamma(h) +  \mathcal{E}^{1+\gamma}(f) \star X_\mu(h),
    \label{tausigleibniz}
\end{equation}
and one has $(X.z)(f):=X(f)\star z=z\star X(f)=(z\cdot X)(f)$, for any $f\in\mathcal{M}_\kappa$ and any $z\in \mathcal{Z}(\mathcal{M}_\kappa)$, the center of $\mathcal{M}_\kappa$. Therefore, $\mathfrak{D}_\gamma$ inherits a structure of $\mathcal{Z}(\mathcal{M}_\kappa^d)$-bimodule. It is worth pointing out that the above derivations are {\it{not}} real derivations, since one can easily verify that
\begin{equation}
    (X_\mu(f))^\dag = -\mathcal{E}^{-2\gamma-1} (X_\mu(f^\dag))
    \label{xpasreel}
\end{equation}
for any $f\in\mathcal{M}_\kappa$.

\paragraph{}
From $\mathfrak{D}_\gamma$, a mere adaptation of the derivation-based differential calculus to incorporate twists gives rise to a family of twisted differential calculus indexed by $\gamma$. The corresponding construction is given in \cite{MW20}. One finds that the differential calculus based on the Lie algebra of twisted derivations $\mathfrak{D}_\gamma$ is completely characterized by the following graded differential algebra
\begin{equation}
    \left(\Omega^\bullet=\bigoplus_{n=0}^{d}\Omega^n(\mathfrak{D}_\gamma),\times,{\bf{d}}\right),\label{diff-algebra}
\end{equation}
where $\Omega^0(\mathfrak{D}_\gamma)=\mathcal{M}_\kappa$, $\Omega^n(\mathfrak{D}_\gamma)$ denotes the linear space of $n$-linear antisymmetric forms, says $\alpha:\mathfrak{D}^{n}_\gamma\to\mathcal{M}_\kappa$ with  $\alpha(X_1,X_2,...,X_n)\in\mathcal{M}_\kappa$ and $\alpha(X_1,X_2,...,X_n\cdot z)=\alpha(X_1,X_2,...,X_n)\star z$, for any $z\in\mathcal{Z}(\mathcal{M}_\kappa^d)$ and any $X_1, \hdots, X_n\in\mathfrak{D}_\gamma$. Note here that the linearity holds true w.r.t. $\mathcal{Z}(\mathcal{M}_\kappa^d)$.

The associative product $\times: \Omega^m(\mathfrak{D}_\gamma) \otimes \Omega^n(\mathfrak{D}_\gamma) \to \Omega^{m+n}(\mathfrak{D}_\gamma)$  is given by
\begin{align*}
    (\omega\times\eta) & (X_1, \dots, X_{m+n}) \\
    &= \frac{1}{m!n!} \sum_{s \in \mathfrak{S}(m+n)} (-1)^{\mathrm{sign}(s)} \omega(X_{s(1)}, \dots, X_{s(m)}) \star \eta(X_{s(m+1)}, \dots ,X_{s(n)})
    \label{ncwedge1}.
\end{align*}
Here, $\mathfrak{S}(m+n)$ is the symmetric group of $m+n$ elements, $\mathrm{sign}(s)$ is the signature of the permutation $s$. One can check that   $\omega \times \eta \ne (-1)^{\delta(\omega) \delta(\eta)} \eta \times \omega$ showing that
$\Omega^\bullet$ \eqref{diff-algebra} is not graded commutative. 

The differential ${\bf{d}}:\Omega^m(\mathfrak{D}_\gamma)\to\Omega^{m+1}(\mathfrak{D}_\gamma)$ is
\begin{equation}
    \left( {\bf{d}}\omega \right) \left( X_1, X_2, \dots, X_{p+1} \right)
    = \sum_{i=1}^{p+1} (-1)^{i+1} X_i \left( \omega(X_1, \dots, \vee_i, \dots, X_{p+1}) \right), 
    \label{ncwedge2}
\end{equation}
for any $m={0},1, \dots, {\left(d-1\right)}$. Here, the symbol $\vee_i$ indicates that the derivation $X_i$ is omitted. One can check that ${\bf{d}}$ satisfies ${\bf{d}}^2=0$ and the twisted Leibniz rule
\begin{equation}
    {\bf{d}}(\omega\times\eta) = {\bf{d}}\omega \times \mathcal{E}^\gamma(\eta) + (-1)^{\delta(\omega)} \mathcal{E}^{1+\gamma}(\omega) \times {\bf{d}}\eta,
\label{leibniz-form}
\end{equation}
in which $\delta({\omega})$ is the degree of $\omega$\footnote{
Note that $\mathcal{E}^x(\omega)\in\Omega^n(\mathfrak{D}_\gamma)$ is defined by $\mathcal{E}^{x}(\omega)(X_1, \dots, X_n)=\mathcal{E}^x(\omega(X_1, \dots, X_n))$
for any $\omega\in\Omega^n(\mathfrak{D}_\gamma)$ and any $X_1, \dots, X_n\in\mathfrak{D}_\gamma$.}.

\paragraph{}
We close this subsection by noticing, first, that the twists involved in the above differential calculi, which are clearly automorphisms of the algebra $\mathcal{M}_\kappa$, are {\bf{not}} $^*$-automorphisms but regular automorphisms. Indeed, one has
\begin{equation}
    (\mathcal{E}^\gamma(a))^\dag = \mathcal{E}^{{-\gamma}}(a^\dag) 
    \ne \mathcal{E}^{\gamma}(a^\dag)
    \label{regular-E}.
\end{equation}

In order to establish a formal bridge between the present situation and the one related to twisted spectral triples briefly discussed at the end of Subsection \ref{section31}, we recall that these latter have regular automorphisms as associated twists. One technical reason (among others) to require this property is that the algebra of the triple must accommodate an action of the modular group. The latter is a one-parameter group of $^*$-automorphisms, says $\{\rho_{t}\}_{t\in\mathbb{R}}$, admitting an analytic continuation $\rho_i$. The modular group is known to be related to {\it{positive}} linear maps $\varphi:\mathcal{A}\to\mathbb{C}$, called KMS weights. These fulfill in particular $\varphi\circ\rho_z=\varphi,\ z\in\mathbb{C}$, where $\rho_z$ is the analytic continuation of $\rho_t$ and $\varphi(a^\dag\star a) = \varphi(\rho_{i/2}(a) \star (\rho_{i/2}(a))^\dag$ for any $a\in\mathcal{A}$ (plus other technical conditions). 

Back to the present situation, the modular group is generated by
\begin{equation}
    \rho_t(f)=e^{\mathrm{i}t\frac{P_0}{\kappa}}(f)=\mathcal{E}^{-\mathrm{i}t},
    \label{sigmat-modul}
\end{equation}
while simple algebraic manipulations yield to the conclusion that the twisted trace given by \eqref{twisted-trace} actually defines a KMS weight $\varphi(a)=\int d^dx\ a(x)$ for any $a\in\mathcal{M}_\kappa$. Finally, the role of the twist $\rho$ of a twisted spectral triple is obviously played by $\mathcal{E}$, while, further assuming $\gamma=0$, the differential $\bf{d}$ defined in \eqref{ncwedge2} plays the role of the differential $\delta_\rho$.

It could be interesting to examine if there is a spectral triple related to the twisted differential calculus presented in this subsection, at least in the $\gamma=0$ case. This latter, if it exists, should presumably bear some similarity with the spectral triple introduced in \cite{Matas}.

\subsection{Connection and curvature.}
\paragraph{}
In \cite{MW20}, \cite{MW20-bis}, a suitable notion of twisted connection has been introduced, which generalizes the, by now, standard notion of noncommutative connection on a right module over the algebra. This is, in essence, the assumption \ref{it:a3}, one of the technical assumptions underlying the present construction listed at the beginning of Subsection \ref{section31}.

The twisted connection is defined as a map $\nabla_{X_\mu}:\mathbb{E}\to\mathbb{E}$ for any $X_\mu\in\mathfrak{D}_\gamma$ such that
\begin{align}
    \begin{split}
        \nabla_{X_\mu+X^\prime_\mu}(m) &= \nabla_{X_\mu}(m) + \nabla_{X^\prime_\mu}(m), \\
        \nabla_{z.X_\mu}(m) &= \nabla_{X_\mu}(m) \star z, \\
        \nabla_{X_{\mu}}(m \star  a) &= \nabla_{X_{\mu}}(m) \star \mathcal{E}^{\gamma}(a) + \mathcal{E}^{\gamma+1}(m) \star X_{\mu}(a),
    \end{split}
    \label{twist-conn}
\end{align}
for any $m\in\mathbb{E}\simeq\mathcal{M}_\kappa$, $z\in\mathcal{Z}(\mathcal{M}_\kappa)$ and $a\in\mathcal{M}_\kappa$.

The corresponding curvature map $\mathcal{F}(X_\mu,X_\nu) := \mathcal{F}_{\mu\nu}: \mathbb{E} \to \mathbb{E}$ is defined by
\begin{equation}
    \mathcal{F}_{\mu\nu} 
    = \mathcal{E}^{1-\gamma} (\nabla_\mu \mathcal{E}^{-1-\gamma} \nabla_\nu - \nabla_\nu \mathcal{E}^{-1-\gamma} \nabla_\mu)
    \label{twist-curvat},
\end{equation}
and fulfills $\mathcal{F}_{\mu\nu}(m\star a) = \mathcal{F}_{\mu\nu}(m)\star a$ as a morphism of module.

Now define
\begin{align}
  \nabla_\mu &:= \nabla_{X_\mu}, & 
  A_\mu &:= \nabla_{\mu}(\mathds{1}), &
  \mathcal{F}_{\mu\nu}(\mathds{1}) &:= F_{\mu\nu}.
  \label{lesdefs}
\end{align}
One easily find that
\begin{equation}
    F_{\mu\nu}
    = \mathcal{E}^{-2\gamma} (X_\mu A_\nu - X_\nu A_\mu)
    + \mathcal{E}^{1-\gamma}(A_\mu) \star \mathcal{E}^{-\gamma}(A_\nu)
    - \mathcal{E}^{1-\gamma}(A_\nu) \star \mathcal{E}^{-\gamma}(A_\mu).
    \label{zecourbure}
\end{equation}
The map $\nabla_{X_\mu}$ defined above extends to a map $\nabla: \mathbb{E}\to\mathbb{E} \otimes \Omega^1(\mathfrak{D}_\gamma)$ with
\begin{align}
    \nabla(a) &= A \star \mathcal{E}^{\gamma}(a) + 1 \otimes {\bf{d}}a, &
    A &\in \Omega^1(\mathfrak{D}_\gamma)
    \label{nabla-diff}
\end{align}
with $A(X_\mu)=A_\mu$ while the map $\mathcal{F}_{\mu\nu}: \mathbb{E} \to \mathbb{E}$ \eqref{twist-curvat} extends to $F: \mathbb{E} \to \mathbb{E} \otimes \Omega^2(\mathfrak{D}_\gamma)$ such that
\begin{equation}
    F = \mathcal{E}^{-2\gamma} ({\bf{d}}A) + \mathcal{E}^{-\gamma} (\mathcal{E}(A) \times A)
    \label{curvat-diff}.
\end{equation}
Furthermore, one can verify the following Bianchi identity:
\begin{equation}
    {\bf{d}}F = \mathcal{E}^{1+\gamma}(F) \times  A - \mathcal{E}^{2}(A) \times \mathcal{E}^{\gamma}(F)
    \label{bianchi}.
\end{equation}

\paragraph{}
According to the assumption \ref{it:a4}, $\mathbb{E}$ is equipped with a Hermitian structure, a sesquilinear form $h:\mathbb{E}\otimes\mathbb{E}\to\mathcal{M}_\kappa$ with $h(m_1,m_2)^\dag=h(m_2,m_1)$ and $h(m_1\star a_1,m_2\star a_2)=a_1^\dag\star h(m_1,m_2)\star a_2$ for any $m_1,m_2\in\mathbb{E}$ and any $a_1,a_2\in\mathcal{M}_\kappa$. Working with a Hermitian module is actually a physically natural requirement. Here, we pick 
\begin{equation}
    h(m_1,m_2)=m_1^\dag \star m_2
    \label{gabuzo}.
\end{equation}

Then, defining as usual the gauge group from the set of automorphisms of $\mathbb{E}$ compatible with the Hermitian structure $h$ given by \eqref{gabuzo}, it is straightforward to verify that the gauge group is entirely characterized by the group of the unitary elements of $\mathbb{E}$ as given by $\mathcal{U}$ \eqref{gauggroup}.

The twisted gauge transformations are \cite{MW20}
\begin{equation}
    \nabla_{{\mu}}^g(a) = (\mathcal{E}^{\gamma+1} \triangleright g^\dag) \star \nabla_{{\mu}}(g \star a),
    \label{gaugebitwist1}
\end{equation}
leading to
\begin{align}
    A_\mu^g &= 
    (\mathcal{E}^{\gamma+1} \triangleright g^\dag) \star A_\mu \star (\mathcal{E}^{\gamma} \triangleright g) 
    + (\mathcal{E}^{\gamma+1} \triangleright g^\dag) \star X_\mu(g), &
    F_{\mu\nu}^g &= \mathcal{E}^2(g^\dag) \star F_{\mu\nu} \star g,
\label{amugtwist}
\end{align}
for any $g\in\mathcal{U}$ and $a\in\mathcal{M}_\kappa$. Notice that the gauge transformation\footnote{
The gauge transformation for the $1$-form connection reads $A^g = (\mathcal{E}^{\gamma+1} \triangleright g^\dag) \times A \star (\mathcal{E}^{\gamma} \triangleright g) + (\mathcal{E}^{\gamma+1} \triangleright g^\dag) \times {\bf{d}}g$. For the 2-form curvature, one has $F^g = \mathcal{E}^2(g^\dag) \times F\times g$.} 
for the curvature {\bf{does not depend on $\gamma$}}. Observe also that $A_\mu$ \eqref{lesdefs} obeys 
\begin{equation}
    A_\mu=\mathcal{E}^{2\gamma+1}\triangleright A_\mu^\dag
    \label{hermit1}
\end{equation}
instead of the usual relation $A_\mu = A_\mu^\dag$ which stems from a twisted condition defining (twisted) Hermitian connections
\begin{equation}
    h(\mathcal{E}^{-1} \triangleright \nabla_{X_{\mu}}(m_1), m_2) 
    + h(\mathcal{E}^{-1} \triangleright m_1, \nabla_{X_{\mu}}(m_2)) 
    = X_\mu h(m_1,m_2)
    \label{cledag}
\end{equation}
for any $X_\mu\in\mathfrak{D}_\gamma$, $m_1,m_2\in\mathbb{E}$.

\subsection{The invariant action.}
\paragraph{}
We are now in position to obtain an action which is a polynomial in the curvature $F_{\mu\nu}$ \eqref{zecourbure} and satisfies the assumptions \ref{it:a1} and \ref{it:a2}, namely being both invariant under $\mathcal{P}_\kappa$ and $\mathcal{U}$ with physically acceptable commutative limit. Let us pick
\begin{equation}
   S_\kappa(F_{\mu\nu})=\int d^dx\ F_{\mu\nu}\star F_{\mu\nu}^\dag,
\end{equation}
and illustrate below the different steps of the computation of the gauge variation of $S_\kappa$.
\begin{align*}
    S_\kappa(F^g_{\mu\nu})
    &= \int d^dx\ \mathcal{E}^2(g^\dag) \star F_{\mu\nu} \star g \star g^\dag \star F_{\mu\nu}^\dag \star \mathcal{E}^{-2}(g) \\
    &= \int d^dx\ \mathcal{E}^2(g^\dag) \star F_{\mu\nu} \star F_{\mu\nu}^\dag \star \mathcal{E}^{-2}(g) \\
    &= \int d^dx\ {\mathcal{E}^{d-1-2}(g) \star \mathcal{E}^2(g^\dag)} \star F_{\mu\nu} \star F_{\mu\nu}^\dag \\
    &=  \int d^dx\ F_{\mu\nu} \star F_{\mu\nu}^\dag
\end{align*}
where, in the second line, we used the definition of $\mathcal{U}$ and the relation for the twisted trace \eqref{twisted-trace} together with \eqref{regular-E} in the third line. From this, it follows that $S_\kappa$ is invariant under the gauge transformations $\mathcal{U}$ whenever
\begin{equation}
    \mathcal{E}^{d-1-2}(g) \star \mathcal{E}^2(g^\dag) = \mathds{1},
\end{equation}
which is satisfied provided
\begin{equation}
    d=5.
\end{equation}

\paragraph{}
Therefore, a salient prediction from this noncommutative gauge theory model is the existence of one extra dimension, to be discussed in Section \ref{section4}.

\subsection{Twisted BRST symmetry.}
\paragraph{}
In this subsection, we assume $\gamma=0$, which does not alter the overall conclusions. As discussed in detail in \cite{brst-twist}, the BRST symmetry leaving invariant the classical action functional is defined by the following operations
\begin{align}
    \begin{split}
        s_0A &= -{\bf{d}}C - A \times C - \mathcal{E}(C) \times A, \\ 
        s_0C &= -C\times C,
    \end{split}
    \label{cestbrst}
\end{align}
where $C$ is the ghost field and the Slavnov operation $s_0$ verifies $s_0^2=0$ and acts as a derivation on bi-graded forms, namely $s_0(\alpha \times \beta) = s_0(\alpha) \times \beta + (-1)^{|\alpha|} \alpha \times s_0(\beta)$ where $|\alpha| := \delta(\alpha) + \mathrm{g}(\alpha)$ is the total degree of the bi-graded form $\alpha$ defined as the sum (modulo 2) of the usual form degree $\delta(\alpha)$ and the ghost number of $\alpha$ denoted by $\mathrm{g}(\alpha)$. This is nothing but the usual algebraic framework related to the BRST symmetry stemming from the very definition of the BRST algebra.

In the commutative case, recall that this latter is a differential algebra built from two copies of the well known Weil algebra \cite{GHV73} with the following structure
\begin{equation}
    \mathcal{W}_{\mathrm{BRST}}(\mathfrak{g})
    = \left( \mathcal{W}(\mathfrak{g}) \otimes \mathcal{W}_{\phi\pi}(\mathfrak{g}),
    \; \widetilde{\bf{d}} = {\bf{d}} + s,
    \; \widetilde{\omega} = A + C,
    \; \widetilde{\Omega} = \widetilde{\bf{d}} \widetilde{\omega} + \frac{1}{2}[\widetilde{\omega}, \widetilde{\omega}] \right),
    \label{BRS-standard} 
\end{equation}
where $(\mathcal{W}(\mathfrak{g}),\bf{d})$ is the Weil algebra\footnote{
This should be supplemented with the Cartan operations \cite{cartan} for $\mathfrak{g}$ which however are not essential for the present summary. } 
of the Lie algebra $\mathfrak{g}$ related to the gauge group with differential $\bf{d}$ \cite{brst-twist}, \cite{GHV73} while $(\mathcal{W}_{\phi\pi}(\mathfrak{g}),s)$ is a copy of $\mathcal{W}(\mathfrak{g})$ now generated, as a free algebra, by elements $C^a$ and $\phi^a$, $a=1,2,...,\text{dim}(\mathfrak{g})$ carrying a new degree, respectively $1$ and $2$, to be called ghost number and the differential $s$. These are the "mirrors" of the generators of the Weil algebra generated by $A^a$ and $F^a$, respectively with form degree  $1$ and $2$, with $F=dA+\frac{1}{2}[A,A ]$ and $A,F\in\mathfrak{g}\otimes \mathcal{W}(\mathfrak{g})$. Indeed, one has $\phi=sC+\frac{1}{2}[C,C]$ and $C, \phi\in\mathfrak{g}\otimes \mathcal{W}_{\phi\pi}(\mathfrak{g})$. The appearance of a bi-graduation of forms introduced just above should now be obvious from \eqref{BRS-standard}.

\paragraph{}
At this stage, is worth recalling that $\widetilde{\bf{d}}={\bf{d}}+s$ is actually the differential for $\mathcal{W}_{\mathrm{BRST}}(\mathfrak{g})$ and that the structure equations which define the BRST symmetry are completely encoded into the famous "Russian formula" \cite{Stora84}
\begin{equation}
  {\widetilde{\Omega}={\Omega}}.  
\end{equation}

It is needless to say that the BRST symmetry is one of the cornerstone of (commutative) Yang-Mills theory and played an essential role in the algebraic theory of perturbative anomalies linked with the Wess-Zumino Consistency Condition whose relevant solution is obtained by solving the $s$-cohomology modulo $\bf{d}$, the higher order cocycles appearing in the corresponding descent equations being associated with a tower of anomalous correlation functions of the BRST current algebra \cite{Stora84, Zumi84, bonora, wal87, dbv-tv}. Note also the essential role played by the BRST symmetry in Topological Field Theories of cohomological class. In the same way, the BRST symmetry is essential in Topological Field Theories of cohomological type \cite{rakow}, \cite{STW94}.

\paragraph{}
Back to the noncommutative case, it turns out that the structure equations \eqref{cestbrst} for a BRST symmetry leaving invariant the classical action functional, $s_0S_\kappa=0$, do not fit within the standard BRST framework sketched just above. In particular, \eqref{cestbrst} does not follow from a (noncommutative analog of) a Russian formula. Moreover, it could be tempting to define $\widehat{d}={\bf{d}}+s_0$ as the suitable differential of the BRST algebra. But this does not make sense as $ \widehat{d}$ does not obey a Leibnitz rule and this triggers problems in defining a suitable Bianchi identity so that the counterpart of $\widetilde{\Omega}$ in \eqref{BRS-standard} can no longer be interpreted as a curvature. This comes from the fact that ${\bf{d}}$ is a twisted differential while $s_0$ is not.

As shown in \cite{brst-twist}, a suitable twisted analog of the BRST algebra is given by
\begin{equation}
    \left( \widehat{\mathbb{W}}, \;
    {\widehat{{\bf{d}}}}_1 = {{\bf{d}}+s_1}, \;
    \widehat{A} = A + C, \;
    \widehat{F} = {\widehat{{\bf{d}}}}_1\widehat{A} + \frac{1}{2}\langle \widehat{A}, \widehat{A} \rangle \right),
    \label{decadix11}
\end{equation}
where $\widehat{\mathbb{W}}$ is essentially defined \cite{brst-twist} in a way similar to the free algebra in \eqref{BRS-standard} and $\widehat{{\bf{d}}}_1$ is now the differential where in particular $s_1$ is twisted with the same twists as $\bf{d}$. The symbol $\langle \cdot, \cdot \rangle$ denotes a graded twisted commutator whose expression is not usefull here (see \cite{brst-twist}). The structure equations defining $s_1$ follow from the Russian formula
\begin{equation}
   \widehat{F} = F 
\end{equation}
leading to
\begin{align}
    \begin{split}
        s_{1}A &= {\bf{d}}C - \mathcal{E}(C) \times A - \mathcal{E}(A) \times C, \\
        s_{1}C &= -\mathcal{E}(C) \times C
    \end{split}
    \label{BRS-algebraic}
\end{align}
with $s_1^2=0$. However, $s_1S_\kappa\ne0$.

Although $s_1$ is no longer a symmetry of the classical action functional $S_\kappa$, it can be shown to be rigidly linked to the actual BRST symmetry of $S_\kappa$ by a continuous transformation. See \cite{brst-twist} for a complete characterization.

\paragraph{}
As far as field theories are concerned, the relevant BRST symmetry is $s_0$ which will serve to generate the useful functional Slavnov-Taylor identity as well as to construct a gauge-fixed action using the standard BRST liturgy.

\vfill\eject

\section{Discussion.}
\label{section4}

\paragraph{}
Let us summarize the main components of the present construction.

\paragraph{}
It should be clear that this latter construction exploits in part some of the salient properties of the convolution algebra for the Lie group related to the Lie algebra of coordinates. At this early stage of the construction, almost all is already fixed. The only freedom one has is the choice of the Haar measure, the right one being chosen in order to insure $\kappa$-Poincar\'e invariance of the action functionals. Then, the convolution and the natural involution equipping the convolution algebra are (simply Fourier) transformed upon quantization by the Weyl map to their (Fourier) counterparts, \textit{i.e.} a star-product and an involution endowing the algebra modeling the $\kappa$-Minkowski space. We note that it should be worth examining the possible impact of this involution in reasonable candidates for analogs of the CPT symmetry.

\paragraph{}
We found that $\kappa$-Poincar\'e invariant gauge theories on $\kappa$-Minkowski space with physically acceptable commutative limit must be 5-d. The gauge invariance
requirement of the action fixes the dimension of the  $\kappa$-Minkowski space to be $d=5$ and selects the unique twisted differential calculus with which the construction can be achieved. Notice by the way that the inclusion of fermions in the present construction is straightforwardly done using the following "covariant derivative" (in obvious notations)
\begin{equation}
    \nabla_\mu \psi = A_\mu \star \mathcal{E}^\gamma(\psi) + X_\mu(\psi).
\end{equation}
It is time to discuss the assumptions used for the present construction which have been given in Subsection \ref{section31}.

\paragraph{}
The first two \ref{it:a1}, \ref{it:a2} are motivated by physical considerations and appear to be quite reasonable. In particular, $\kappa$-Poincar\'e invariance is a rather natural assumption together with the need to recover an usual gauge theory, here QED, at the commutative $\kappa\to\infty$ limit. In more physical words, one considers systems where the ((quantum) space and gauge) symmetries together with the gauge invariant action vary along with the energy scale.

\paragraph{}
The assumption \ref{it:a4} in some sense serves to work with a noncommutative analog of a Yang-Mills connection thanks in particular to the hermiticity condition of the connection which arises through the introduction of a Hermitian structure. In this respect, \ref{it:a4} can be viewed as supplementing \ref{it:a1} and \ref{it:a2}. The choice $\mathbb{E}\simeq\mathcal{M}_\kappa$ triggers a noncommutative analog of a $U(1)$ gauge theory and is also convenient as it simplifies the analysis: only $U(1)$ gauge theory is considered, which is enough for the moment. Generalisation to noncommutative analogs of $U(n)$ gauge theories can be obtained by starting from a module built from the direct product of $n$ copies of $\mathcal{M}_\kappa$, \textit{i.e.} $\mathbb{E} \simeq \mathcal{M}_\kappa^{n}$. We do not expect that new features should appear compared to the case reported here.

\paragraph{}
Let us comment the assumption \ref{it:a5}. One may wonder if the conclusion obtained about the distinguished value of the dimension of the $\kappa$-Minkowski space $d=5$, for which both $\mathcal{P}_\kappa$ and gauge $\mathcal{U}$ invariances can coexist, survives if an additional twist is included. In fact, twisting the action of the algebra on the module leads to \eqref{zeaction} in which now $\sigma\ne\text{Id}$, $\sigma\in\text{Aut}(\mathcal{M}_\kappa)$. As shown in details in \cite{module-paper}, this only modifies the hermiticity condition related to the gauge potential $A_\mu$ as
\begin{equation}
    \sigma^{-1}(A_\mu) = \mathcal{E}^{2\gamma+1}(\sigma^{-1}(A_\mu)^\dag)
    \label{twistedhermit},
\end{equation}
which boils down to $A_\mu = \mathcal{E}^{2\gamma+1}(A_\mu^\dag)$ or $A_\mu = \mathcal{E}^{2\gamma+1}\sigma^2(A_\mu^\dag) $ whenever $\sigma$ is a $^*$-automorphism or a regular automorphism respectively, which commutes with $\mathcal{E}$. For a general $\sigma$, the gauge group $\mathcal{U}$ is essentially unchanged. The same conclusion holds true for the distinguished value for the dimeension of $\mathcal{M}_\kappa$. One still ends up with $d=5$.

\paragraph{}
Hence, a physically salient prediction for these $\kappa$-Poincar\'e invariant gauge theories on $\kappa$-Minkowski space is the existence of one extra-dimension. Phenomenological properties stemming from this extra dimension have been explored and discussed in \cite{MW20-bis} in the framework of models with Universal Extra Dimension \cite{UED}. Whenever the extra dimension is compactified on the simple orbifold $\mathbb{S}^1/\mathbb{Z}_2$, it turns out that consistency with the data from LHC constraining the size $\mu^{-1}$ of the extra dimension yields $\mu\gtrsim\mathcal{O}(1-5)\ \text{TeV}$. This leads to
\begin{equation}
\kappa\gtrsim\mathcal{O}(10^{13})\ \text{GeV}\label{kappaconserv1},
\end{equation}
where $\kappa$ in \eqref{kappaconserv1} is identified with the 5-dimensional bulk Planck mass.

Observational constraints from Gamma Ray Bursts (GRB) \cite{zereview} may be used to improve this bound \cite{MW20}. Indeed, the in-vacuo dispersion relation for the (4-d) photon (actually the zero-mode of the 5-dimensional $A_\mu$ field in the compactification scheme mentionned above) is entirely fixed by the noncommutative differential calculus together with the kinetic operator for $A_\mu$. Expanding this dispersion relation in inverse powers of $\kappa$ yields
\begin{equation}
    E^2-|\vec{p}| ^2-\frac{1}{\kappa}E^3+\mathcal{O}(\frac{1}{\kappa^2}) = 0
    \label{dispersion}.
\end{equation}
From recent data on GRB, one infers that
\begin{equation}
    \kappa\gtrsim\mathcal{O}(10^{17}-10^{18})\ \text{GeV}
    \label{kappa-mgm}.
\end{equation}

Now, within the scenario of UED models, the Planck mass $M_p$, $\kappa$ and $\mu$ are related together by
\begin{equation}
    M_P^2=\frac{\kappa^3}{\mu} ,
\end{equation}
which combined with \eqref{kappa-mgm} leads to
\begin{equation}
    \mu\gtrsim\mathcal{O}(10^{13}-10^{16})\ \text{GeV},
\end{equation}
thus corresponding to a very small size for the extra dimension within the scenario for UED models.

\paragraph{}
The BRST machinery elaborated in \cite{brst-twist} can be used to explore perturbative properties of the gauge theory model described by $S_\kappa$. This has been initiated in \cite{tadpole} where it has been found that a non-vanishing one-point function shows up at the one-loop order. This leads to a contribution linear in $A_\mu$ in the one-loop effective action of the form
\begin{equation}
    \Gamma^1(A)=\int d^5x \ K(\kappa)A_0(x),\label{gamma1}
\end{equation}
where $K(\kappa)$ is a diverging integral (which can be regularized). Note that the computation has been carried out using a kind of noncommutative generalization of the Lorentz gauge, i.e. $X_\mu A_\mu=0$. Non-vanishing tadpole in other noncommutative gauge theories are already known to occur, as for instance the massless gauge theory on $\mathbb{R}^3_\lambda$ \cite{tadpole-spatial}, the matrix gauge models on the 2-d Moyal space $\mathbb{R}^2_\theta$ \cite{vign-sym}. Note that the inherent structure of these matrix gauge models renders the investigation of the 4-dimensional case \cite{wal-moyal1} prohibitively complicated. Recall however that no tadpole occurs in a class of widely studied gauge theories on $\mathbb{R}^4_\theta$ \cite{Matu}. The contribution \eqref{gamma1} indicates that 
the classical vacuum of the theory is not stable against quantum
fluctuations. Note that \eqref{gamma1} is not gauge (BRST) invariant, suggesting that the classical symmetry is broken.\\

\section{Outlook.}
\label{sec:outlook}

\paragraph{}
Through their gauge theoretic formulation, noncommutative (quantum) spaces may also involve gravity approach. This path is though to lead to an approach to quantum gravity theory and has sparked a huge literature. We review here the main theoretical framework to incorporate gravity in a noncommutative framework.

\paragraph{}
For derivation based differential calculus over an algebra $\mathcal{A}$, the noncommutative analog of vector fields are derivations, $\mathrm{Der}(\mathcal{A})$. It follows that one would like to promote $\mathrm{Der}(\mathcal{A})$ to a module over the algebra, that is, with our notations assumption \ref{it:a4} becomes $\mathbb{E} \simeq \mathrm{Der}(\mathcal{A})$. However, $\mathrm{Der}(\mathcal{A})$ is not a $\mathcal{A}$-module but a $\mathcal{Z(A)}$-module. But defining a connection on a $\mathcal{Z(A)}$-module would not capture in some sense the noncommutative essence of a gravity theory.

\paragraph{}
A first way out is provoded by the framework of central bimodules \cite{Mourad_1995, Dubois_Violette_1996} in which one can uniquely associate a connection on the $\mathcal{A}$-module $\Omega^1(\mathcal{A})$, the set of one-forms on $\mathcal{A}$, to a connection on $\mathrm{Der}(\mathcal{A})$. A  correspondence between $\mathcal{A}$-modules and $\mathcal{Z(A)}$-modules, seen as duals of each other, is made in \cite{Dubois_Violette_1996}. Furthermore, a metric is defined as a non-degenerate symmetric $\mathcal{Z(A)}$-bilinear complex map on $\mathrm{Der}(\mathcal{A}) \times \mathrm{Der}(\mathcal{A})$ and thus a Levi-Civita connection is defined and found to be unique. 

Central bimodules have generated various direct applications as in  \cite{Dabrowski_1996, Madore_1995,Dubois_Violette_1995, twoparam-lin}. Other frameworks listed below make use of central bimodules.

\paragraph{}
One of these frameworks, which can be viewed as a second way out, is based on tame differential calculus \cite{Bhowmick_2020a, Bhowmick_2020b}. According to these works, the set of one-forms $\Omega^1(\mathcal{A})$ of a differential calculus, that is tame, is a central $\mathcal{A}$-module, a notion defined by the authors, which extends the notion of centered bimodule. A metric is defined as a non-degenerate symmetric $\mathcal{A}$-bilinear complex map on $\Omega^1(\mathcal{A}) \times \Omega^1(\mathcal{A})$. A Levi-Civita connection is found to exists and to be unique.

In this context, the noncommutative version of vector fields is defined through the metric and is isomorphic to $\mathrm{Der}(\mathcal{A})$ for a tame differential calculus based on derivations. This approach comes from an earlier work \cite{Bhowmick_2019}.

\paragraph{}
A third recent way out is based on braided geometry \cite{Aschieri_2020}. This study considers a less wide class of algebra which are $\mathcal{H}$-module braided symmetric, or braided commutative, algebras $\mathcal{A}$, for $\mathcal{H}$ a {{triangular}} Hopf algebra. $\mathcal{A}$ plays the role of the base space and $\mathcal{H}$ its symmetries. Using braided symmetry, the set of braided derivations, that is derivations satisfying a braided Leibniz rule, is found to be a $\mathcal{A}$-module. For details and proofs on this part see for example \cite{Weber_2020_PhD}. Finally, two kinds of connections emerge: left braided and right braided. A metric is defined as a non-degenerate braided symmetric $\mathcal{A}$-bilinear complex map on $\Omega(\mathcal{A}) \otimes \Omega(\mathcal{A})$ and so two unique (one left and one right braided) Levi-Civita connections are found.

The idea to use such structure comes from the attempt of \cite{Weber_2020} to build a noncommutative Cartan calculi that escapes from central bimodules, and so to the center of the algebra, but that sticks to derivation based differential calculus.

The braided geometry was also implemented in a theory of gravity expressed via $L_\infty$ algebras \cite{Ciric_2021}.

This formalism generalizes the one previously studied by \cite{Aschieri_2006} of Drinfeld twists applied to the canonical Hopf algebra of the universal enveloping algebra of vector fields. Such a framework was first used on the Moyal space \cite{Aschieri_2005}. Even if this setting was developed earlier, recent interesting results are derived with it like in \cite{Aschieri_2021a}.

\paragraph{}
Drinfeld twists were also extended to fuzzy spaces \cite{Kurkcuoglu_2006}. To do so the star product was simplified and a "pseudo-twist" was defined. Indeed, fuzzy spaces have been widely studied when it comes to gravity \cite{Madore_1997, Nair_2003, Abe_2003, Valtancoli_2004, Manolakos_2019}. 

\paragraph{}
One of the earliest attempts of developing a noncommutative version of gauge theory was done by \cite{Brzezinski_1993, Hajac_1996} through the notion of quantum principal fiber bundle, also called Hopf-Galois extensions. This object aims at generalizing the notion of principal fiber bundles when the structure group is a quantum group (Hopf algebra) and so to use algebraic versions of all elements of principal bundles. Note that other approaches to a quantum version of the principal fiber bundle exist in the literature like \cite{Pflaum_1994, Durdevic_1996}. This framework rapidly turned to Riemannian geometries \cite{Majid_1997, Beggs_2009}. The idea lies in a new formalism of classical Rieamannian geometry based on frame resolution, generalising the frame bundle. This appraoch was found compatible with Drinfeld twists \cite{Aschieri_2016b, Aschieri_2016a} and was enlarged to non-affine bases with a sheaf theoretic setting \cite{Aschieri_2021b, Aschieri_2021c}.

\paragraph{}
The question of how to incorporate noncommutative features into gravity theory has been pursued through other ways.

One of them relies on Poincar\'e gauge gravity \cite{Utiyama_1956, Kibble_1961, Hehl_1976}, which is a gauge theory based on gauging the Poincar\'e group and produces a spin-torsion theory, also called $U_4$ theory or Einstein-Cartan(-Sciama-Kibble) theory, that is gravity with torsion. The main idea is to gauge a quantum version of the Poincar\'e group, explicitly $ISO_q(1,3)$ \cite{Castellani_1994, Bimonte_1998a, Bimonte_1998b}.

With the growth of interest for the Moyal space, several studies were performed on classical gravity theory turned noncommutative by replacing the usual product by a star product. In these approaches, the question of diffeomorphism invariance is central since space-time noncommutativity breaks the classical version of this invariance. Studies was performed for $U(n)$ gauges \cite{Cacciatori_2002a, Cacciatori_2002b, Deliduman_2006, Balachandran_2006, Alvarez_2006, Chamseddine_2003, Castellani_2013, Aschieri_2014, Cardella_2003} and $GL_n(\mathbb{C})$ gauges \cite{Chamseddine_2004}, but, since complex metrics yields non-physical behaviours \cite{Moffat_2000}, some authors turned to noncommutative gauge theory of $SO$ and $Spin$ groups \cite{Bonora_2000, Jurco_2000, Bars_2001, Dimitrijevic_2012, Dimitrijevic_2014, Moffat_2000, Chamseddine_2001b}. The link between Moyal space and gravity was first made in string theory and a hole field of literature focuses on the two as evoked in the review \cite{Szabo_2006}. 

Another way to make a noncommutative version of gravity would be to have a noncommutative version of the frame field, also called tetrad, vierbein or veilbein. This was first done by complexifying the tetrad and adding star products in every definitions, like in \cite{Moffat_2000, Chamseddine_2001b}. But \cite{Buric_2005} went a step further and fixed the hole definition of their noncommutative geometry on an algebraic version of the tetrad.

\paragraph{}
There were several attempts to make gravity on noncommutative space through the framework of spectral triples. One of the earliest attempt was made by \cite{Chamseddine_1993} and refined by \cite{Landi_1994,Sitarz_1994} on Kaluza-Klein theory. They particularly showed that their notion of metric as a non-degenerate Hermitian inner product coincides with the noncommutative distance defined by \cite{Connes_1991}. A more recent version of this study is made in \cite{Chamseddine_2015}.

Simultaneously, \cite{Connes_2014} and \cite{Fathizadeh_2013} used the spectral triple formalism to compute the curvature of the noncommutative dilated two-torus. The latter is defined as the second term of the heat kernel expansion, in analogy with classical Riemannian geometry. This study was followed by many others on the subject, like \cite{Fathizadeh_2015, Lesch_2016}. Following this path and inspired by the work of \cite{Rosenberg_2013}, which was revised in \cite{Peterka_2017}, \cite{Arnlind_2017} defined a version of pseudo-Riemannian calculi of modules over noncommutative algebras.

\paragraph{}
Some authors \cite{Calmet_2005, Banerjee_2007, Much_2019} tried to constraint gravity on noncommutative spaces by considering general Lie algebraic relations for the space-time noncommutativity.

\paragraph{}
Finally, there are various other approaches to noncommutative gravity like flat cleft extensions \cite{Majid_2013}, quantum jet bundles \cite{Majid_2022}, self-dual gravity \cite{Garcia_2003, Estrada_2008, Grezia_2014}, teleparallel gravity \cite{Langmann_2001, Nishino_2002}, symplectic gravity \cite{Miao_2011, Lee_2014}, so-called $\kappa$-deformed metrics \cite{Harikumar_2017}, curved Moyal product \cite{Harikumar_2006} or through AdS/CFT correspondence  \cite{Jevicki_1999}.\\

The above list gives a summary of the (numerous) attempts to set up a suitable framework to incorporate gravity.
A suitable generalization of the results summarized in Section 2-4 to encompass gravity should of course modify in some extent the notion of noncommutative connection on a right module which underlies the whole construction. We have undertaken this task and related results will be reported in forthcoming publications.

\paragraph{}
\noindent{\bf{Acknowledgments:}} We thank the organizers of the Corfu Summer Institute and the Workshop on Quantum Geometry, Field Theory and Gravity for their invitation. We also thank the Action CA18108 QG-MM, ”Quantum Gravity
Phenomenology in the multi-messengers approach”, from the European Cooperation in Science and Technology (COST).


\end{document}